\documentclass[%
preprint,
nofootinbib,
 amsmath,amssymb,
 aps,
]{revtex4-1}

\usepackage{graphicx}
\usepackage{dcolumn}
\usepackage{bm}

\pdfoutput=1

\usepackage{amsmath,amssymb,latexsym}

\usepackage{times}

\usepackage{epstopdf}

\usepackage{graphics,epsfig}

\usepackage{color}

\usepackage{mathrsfs}

\DeclareMathAlphabet{\mathpzc}{OT1}{pzc}{m}{it}

\let\a=\alpha \let\b=\beta \let\g=\gamma \let\d=\delta \let\e=\epsilon
\let\z=\zeta  \let\th=\theta  \let\k=\kappa
\let\l=\lambda \let\m=\mu \let\n=\nu \let\x=\xi \let\p=\pi 
\let\s=\sigma   \let\f=\phi  
 
       \let\D=\Delta \let\Th=\Theta 
\let\X=\Xi  \let\S=\Sigma  \let\Y=\Psi
 \let\W=\Omega
\let\la=\label  
  
 \def\bd{\begin{document}} \def\ed{\end{document}}
\def\ds{\documentstyle} \let\fr=\frac \let\bl=\bigl \let\br=\bigr
\let\Br=\Bigr \let\Bl=\Bigl
\let\bm=\bibitem
\let\na=\nabla
\def\tU{{\widetilde U}}
\let\pa=\partial \let\ov=\overline
\def\ie{{\it i.e.\ }}
\newcommand{\be}{\begin{equation}}
\newcommand{\ee}{\end{equation}}
\def\ba{\begin{array}}
\def\ea{\end{array}}
\def\ft#1#2{{\textstyle{{\scriptstyle #1}\over {\scriptstyle #2}}}}
\def\fft#1#2{{#1 \over #2}}
\def\F#1#2{{ F_{#1}^{(#2)} }}
\def\cF#1#2{{ {\cal F}_{#1}^{(#2)} }}

\def\R{{\bf R}}
\def\sst#1{{\scriptscriptstyle #1}}
\def\oneone{\rlap 1\mkern4mu{\rm l}}
\def\e7{E_{7(+7)}}
\def\td{\tilde}
\def\wtd{\widetilde}
\def\im{{\rm i}}
\def\bog{Bogomol'nyi\ }
\newcommand{\ho}[1]{$\, ^{#1}$}
\newcommand{\hoch}[1]{$\, ^{#1}$}
\newcommand{\bea}{\begin{eqnarray}}
\newcommand{\eea}{\end{eqnarray}}
\newcommand{\ra}{\rightarrow}
\newcommand{\lra}{\longrightarrow}
\newcommand{\Lra}{\Leftrightarrow}
\newcommand{\bp}{\tilde \beta^\prime}
\newcommand{\cB}{{\cal B}}
\newcommand{\cO}{{\cal O}}
\newcommand{\vecx}{\vec{x}}
\newcommand{\vecy}{\vec{y}}
\newcommand{\vecp}{\vec{p}}
\newcommand{\vecq}{\vec{q}}
\newcommand{\tr}{{\rm tr} }
\newcommand{\Tr}{{\rm Tr} }
\newcommand{\NP}{Nucl. Phys. }

\newcommand{\cL}{{\cal L}}
\newcommand{\cA}{{\cal A}}
\newcommand{\cT}{{\cal T}}
\newcommand{\cR}{{\cal R}}
\newcommand{\cD}{{\cal D}}
\newcommand{\cH}{{\cal H}}

\def\Cb{\bar{C}}

\def\sst#1{{\scriptscriptstyle #1}}
\def\0{{\sst{(0)}}}
\def\1{{\sst{(1)}}}
\def\2{{\sst{(2)}}}
\def\3{{\sst{(3)}}}
\def\4{{\sst{(4)}}}
\def\5{{\sst{(5)}}}
\def\6{{\sst{(6)}}}
\def\7{{\sst{(7)}}}
\def\8{{\sst{(8)}}}
\def\9{{\sst{(9)}}}
\def\p{{\sst{(p)}}}
\def\q{{\sst{(q)}}}
\def\ve{\varepsilon}
\def\vf{\varphi}
\def\F{\Phi}
\def\wg{\wedge}

\def\thb{\bar{\theta}}
\def\Thb{\bar{\Theta}}
\def\barp{\bar{p}}
\def\barq{\bar{q}}
\def\barc{\bar{c}}
\def\bard{\bar{d}}
\def\e{\epsilon}

\def \bi{\bibitem}
\def \la {\label}

\def \l {\lambda}
\def\foot{\footnote}
\def \tl  {{\tilde \l}}
\def \sql {{\sqrt \l}}
\def \adss {$AdS_5 \times S^5$\ }
\newcommand{\rf}[1]{(\ref{#1})}
\def \ov {\over}

\def\th{\theta}
\def\Th{\Theta}
\def\vth{\vartheta}
\def\btheta{{\bar\theta}}
\def\ttheta{{{\tilde\theta}}}
\def\bttheta{{{\bar\ttheta}}}
\def\vth{\vartheta}

\def\ra{\rightarrow}
\def\N{\nabla}
\def\F{{\cal F}}
\def\uM{\underline{M}}
\def\uA{\underline{A}}
\def\uN{\underline{N}}
\def\uP{\underline{P}}
\def\ua{\underline{a}}
\def\ub{\underline{b}}
\def\uc{\underline{c}}
\def\ud{\underline{d}}
\def\ue{\underline{e}}
\def\uf{\underline{f}}
\def\ui{\underline{i}}
\def\uj{\underline{j}}
\def\uk{\underline{k}}
\def\ul{\underline{l}}
\def\ual{\underline{\alpha}}
\def\ube{\underline{\beta}}
\def\um{\underline{m}}
\def\un{\underline{n}}
\def\up{\underline{p}}
\def\uq{\underline{q}}
\def\ur{\underline{r}}
\def\us{\underline{s}}
\def\umu{\underline{\mu}}
\def\unu{\underline{\nu}}
\def\ula{\underline{\l}}
\def\uka{\underline{\k}}
\def\usi{\underline{\s}}
\def\urh{\underline{\r}}
\def\cc{\circ}
\def\eqv{\equiv}

\def\ni{\noindent}

\def\Ep{E^{{}^{(+)}}}
\def\Em{E^{{}^{(-)}}}

\def\Mp{M^{{}^{(+)}}}
\def\Mm{M^{{}^{(-)}}}

\def \ha{{1\ov 2}}

\def\r{\rho}

\def\Y{{\rm Y}}
\def\X{{\rm X}}
\def\tY{\tilde{\rm Y}}
\def\tX{\tilde{\rm X}}
\def\dY{\dot{\rm Y}}
\def\dX{\dot{\rm X}}

\def \J {\mathcal{J}}
\def \del {\partial}

\def\dF{\dot{F}}
\def\dG{\dot{G}}
\def\df{\dot{f}}
\def \E {{\cal E}}
\def \S {{\cal S}}
\def \J {{\cal J}}

\def\ms{\mathcal{S}}
\def\mj{\mathcal{J}}
\def\soj{\fr{\ms}{\mj}}
\def \R {{\bf R}}
\def \om {\omega}
\def \bE {\bar E}
\def \x {{\cal X}}

\def \bi{\bibitem}
\def \la {\label}

\def \l {\lambda}
\def\foot{\footnote}
\def \tl  {{\tilde \l}}
\def \sql {{\sqrt \l}}
\def \adss {$AdS_5 \times S^5$\ }
\def \ov {\over}

\def \varpi {{\rm w}}

\def\thb{\bar{\theta}}
\def\Thb{\bar{\Theta}}
\def\mb{\bar{\m}}
\def\ab{\bar{\a}}
\def\zb{\bar{z}}
\def\psib{\bar{\psi}}
\def\barp{\bar{p}}
\def\barq{\bar{q}}
\def\barc{\bar{c}}
\def\bard{\bar{d}}
\def\e{\epsilon}
\def\wb{\bar{w}}
\def\lb{\bar{\l}}
\def\Jb{\bar{J}}
\def\Nb{\bar{N}}
\def\Zb{\bar{Z}}
\def\pab{\bar{\pa}}

\def\At{\tilde{A}}
\def\Bt{\tilde{B}}
\def\Ct{\tilde{C}}
\def\Dt{\tilde{D}}
\def\Et{\tilde{E}}
\def\Ft{\tilde{F}}
\def\Gt{\tilde{G}}
\def\Ht{\tilde{H}}
\def\Kt{\tilde{K}}
\def\Mt{\tilde{M}}
\def\Nt{\tilde{N}}
\def\Rt{\tilde{R}}
\def\at{\tilde{a}}
\def\bt{\tilde{b}}
\def\ct{\tilde{c}}
\def\dt{\tilde{d}}
\def\et{\tilde{e}}
\def\ft{\tilde{f}}
\def\htil{\tilde{h}}
\def\gt{\tilde{g}}
\def\nt{\tilde{n}}
\def\mut{\tilde{\mu}}
\def\nut{\tilde{\nu}}
\def\pht{\tilde{\f}}
\def\Pht{\tilde{\Phi}}
\def\vft{\tilde{\vf}}
\def \zet{\tilde{\z}}

\def\rht{\tilde{\rho}}

\def\asth{\hat{*}}
\def\phh{\hat{\phi}}

\def\bA{{\bf A}}

\def\ola{\overleftarrow}
\def\ora{\overrightarrow}
\def\alt{\tilde{\a}}

\def\eh{\hat{e}}
\def\eph{\hat{\e}}
\def\ph{\hat{p}}
\def\alh{\hat{\a}}
\def\beh{\hat{\b}}
\def\gah{\hat{\g}}
\def\Fh{\hat{F}}
\def\muh{\hat{\m}}
\def\nuh{\hat{\n}}
\def\thh{\hat{\th}}
\def\rhh{\hat{\r}}
\def\dh{\hat{d}}
\def\ih{\hat{i}}
\def\jh{\hat{j}}
\def\hh{\hat{h}}
\def\nh{\hat{n}}
\def\gh{\hat{g}}
\def\kh{\hat{k}}
\def\deh{\hat{\d}}
\def\wh{\hat{w}}
\def\lah{\hat{\l}}
\def\Ah{\hat{A}}
\def\Kh{\hat{K}}
\def\Nh{\hat{N}}
\def\Rh{\hat{R}}
\def\Ch{\hat{C}}
\def\Omh{\hat{\Omega}}

\def\xh{\hat{x}}

\def\ps{\rlap{\, /}\;\,p }
\def\ks{\rlap{\, /}\;\,k }

\def\gym{g_{YM}}

\def\adot{\dot{a}}
\def\bdot{\dot{b}}
\def\bpa{\bar{\pa}}

\def\pr{\prime}
\def\ssk{\medskip}
\def\clb{\color{blue}}
\def\clr{\color{red}}
\def\clg{\color{green}}

\def\bfA{{\bf A}}
\def\bfB{{\bf B}}
\def\bfK{{\bf K}}
\def\bfU{{\bf U}}
\def\bfX{{\bf X}}
\def\bfY{{\bf Y}}
\def\bfZ{{\bf Z}}
\def\bfg{{\bf g}}
\def\bfn{{\bf n}}

\def \vk{\vec{k}}
\def \vx{\vec{x}}

\def\bsk{\bigskip}
\def\ssk{\medskip}

\def\Ec{{\cal E}}

\def\hb{\hbar}

\begin{document}


\title{The Stark effect in superfluid $^4$He with relative flows}

\author{A. S. Rybalko}
\email{rybalko@ilt.kharkov.ua}
\author{S. P. Rubets}
\author{E. Ya. Rudavskii}
\affiliation{%
Verkin Institute for Low Temperature Physics and Engineering\\
47 Nauki Av., Kharkov 61102 UA
}%


\author{R. V. Golovashchenko}
\author{S. I. Tarapov}
\author{V. N. Derkach}
\affiliation{
Usikov Institute for Radiophysics and Electronics\\
12 Ak. Proskury, Kharkov 61085 UA 
}%
\author{V. D. Khodusov}
\author{A. S. Naumovets}
\author{A. J. Nurmagambetov}
\email{ajn@kipt.kharkov.ua}
\altaffiliation[Also at Akhiezer Institute for Theoretical Physics of NSC KIPT, 1 Akademicheskaya St., 61108 Kharkov UA \& Usikov Institute for Radiophysics and Electronics,
12 Ak. Proskury, Kharkov 61085 UA.]{}
\affiliation{%
Karazin Kharkov National University\\
4 Svobody Sq., Kharkov 61022 UA 
}%


\date{\today}

\begin{abstract}
We conducted series of experiments on observing a Stark-type effect in superfluid $^4$He in presence of relative laminar flows of the normal and superfluid components. It is designed a measurement cell which allows us to simultaneously create hydrodynamic flows in the liquid and to carry out high-frequency radio-measurements at external electric field. We used a dielectric disk resonator that made possible to cover a wide frequency range. In our experiments it was registered the spectrum of the dielectric disk resonator modes, as well as narrow lines of absorption of a microwave radiation in He II on its background and in different conditions. We discovered that having in the liquid helium a relative motion of the normal and superfluid fractions in the temperature range of 1.4$\div$2.17\,K the narrow line of absorption/radiation is observed in the EM spectrum, the frequency of which -- 180\,GHz -- corresponds to the roton minimum. This line splits in a constant electric field. Note that in a weak electric field the value of splitting depends linearly on the electric field strength, i.e. the linear Stark effect is detected. It is found that with the external electric field increasing both split lines are displaced towards more low frequencies side. The obtained data set could be described by an empirical formula, taking into account as the linear part of the Stark effect, as well as a quadratic addition, related to the polarization part. The data point out on having particles or excitations in the liquid helium with the dipole moment $\sim 10^{-4}$\,D, that in four order less of the characteristic dipole moment of polar molecules. The comparison of our findings to values of the electric dipole moment (EDM) of elementary particles and nuclei is also performed. We sum up with brief discussion of extensions of the known theoretical models and possible mechanisms of the EDM production.

\begin{description}
\item[Keywords]
Superfluidity, Stark effect, electric dipole moment, photons, rotons, recoil impulse. 
\item[PACS numbers]
07.05.Fb; 07.50.--e; 07.57.--c; 32.60.+i; 32.10.Dk; 67.40.--w; 67.40.Bz; 67.90.+z  

\end{description}
\end{abstract}

\pacs{Valid PACS appear here}
\maketitle


\section{\label{sec:level1} Introduction}

Symmetry plays a key role in physics. Invariance under continuous transformation groups underlies modern theories of elementary particle interactions; discrete transformations are involved in fundamental and applied physics, from Cosmology to Spectroscopy of atoms and molecules. All physical processes, described in frameworks of standard (local) QFT, have to obey the prominent Schwinger-L\"uders-Pauli CPT-theorem. Before the Cronin-Fitch discovery of the CP-breaking in Strong Interactions, it was believed in deterministic (upon the evolution) nature of QFT. Currently, breaking the time reversal in processes where the CP-invariance does not hold is commonly admitted. 

Breaking the CP-invariance is considered as one of the most viable scenario to explain the observed mismatch between matter and anti-matter. The Standard Model and its extensions give various sources of the CP-violation; the electric dipole moment of quarks and leptons is among them. The latter came in the focus of our attention in view of unexpected finding the response of the $^{4}$He Bose-Einstein condensate to an external source similar to the reaction of a medium endowed with the electric dipole moment.

Recall, the single atom liquid of $^{4}$He atoms with zero electric dipole moment in a free state is characterized by the absence of the solid state phase with approaching the absolute zero, reveals the second order phase transition similar to that of magnetics and segnetics upon the critical temperature $T=T_\l$, and possesses the superfluidity below the $T_\l$ \cite{Patterman:1974}. Numerous experiments on birefringence \cite{Kikoin:1936,Auzinsh:2004}, as well as on the temperature dependence of dielectric susceptibility (within the range of 1$\div$10\,kHz) \cite{Berthold:1976,Kierstead:1976,Chan:1977,Stankowski:1986}, also confirm the absence of electric moments in the liquid. We can notice it happened due to overlooking the dynamical properties of He II. However, experiments on revealing the influence of electric field on the superfluid liquid velocity in narrow slots \cite{Neidhardt:1962} and on the transport over the superfluid film \cite{Eselson:1974}, as well as the Raman scattering of light \cite{Greytak:1969}, pointed out on possible manifestation of electrohydrodynamical properties at the seemingly electroneutral liquid. In the case of the light scattering, as specified by the superfluid theory and in accordance with the fulfillment of the energy and momentum conservation laws, it was believed that the optical photon disappears and the Stokes photon is created. The difference between frequencies of the excited light and of the resulted Stokes' photon is $\d f=360$\,GHz, so that it corresponds to the threshold energy $2\triangle=h\d f$ ($h$ is the Planck constant) to create two quasiparticles -- rotons -- with the opposite momenta \cite{Greytak:1970}. 

The studies of \cite{Neidhardt:1962,Eselson:1974,Greytak:1969,Greytak:1970} initiated new directions of investigations in electric properties of the He II liquid. Starting from 2004 two series of experiments were performed on measuring the electric properties of the liquid helium, which dramatically changed the view on the single-atom liquid as on a collection of interacting structureless spherical particles.

The result of the first series of tests was consist in discovering the electric response (spontaneous polarization) of He II upon the relative motion of normal and superfluid components in the second sound wave and in the absence of an external electric field \cite{Rybalko:2004,Rybalko:2005,Rybalko:2011,Chagovets:2016,Chagovets:2016pb,Rybalko:2017,Chagovets:2017,Yayama:2018,Natsik:2020}. Herewith, it was observed the sharp difference in the laminar and turbulent flows (see \cite{Natsik:2020}). 

To figure out the origin of the obtained response, we carried out the second series of tests with a dielectric disk resonator (DDR), embodied into the liquid helium \cite{Rybalko:2007,Rybalko:2008,Rybalko:2008-1,Rybalko:2009}. We studied the interaction of the liquid with whispering gallery modes of the DDR in the range, overlapping in frequency the rotonic part of the helium energy spectrum (100$\div$200 GHz). It was shown that the absorption spectrum of the optically transparent quantum liquid has the resonant peculiarities at the frequency $f_{rot}=\triangle/h$; namely, a wide (several GHz) absorption resonance. As in before, $\triangle$ is the energy gap corresponding to the creation of the roton quasiparticle. With approaching the temperature to $T_\l$ the resonance width increased and a deformation occurred that made impossible to extrapolate the resonance by the gaussian. The situation got changed with presence of artificially created relative flows of the helium components around the resonator, when the flows coincide (or being opposite) in direction with the Poynting vector of the DDR whispering gallery wave \cite{Rybalko:2007,Rybalko:2008,Rybalko:2008-1,Rybalko:2009,Rybalko:2010}.  In that case it is observed a narrow resonance in the background of a basement-type wide resonance, the location of which is almost independent on the temperature, and which, in dependence as on the pumping power as well as on the power of relative flows, undergo three stages: the absorption, the electromagnetic transparency, the induced radiation. It became clear that the observed effect is caused by the difference between absorption and induced radiation \cite{Eliashevich:2001}. {\it A contradiction between theory and experiment} consists in the fact that within the existent roton models \cite{Donnelly:1997,Bedell:1982,Galli:1996}  it is impossible to explain as the process of the resonant absorption of EM waves of the frequency 180\,GHz \cite{Rybalko:2007}, as well as the absorption of single photons of the roton frequency, generated by the embedded into He II and further excited dysprosium atoms \cite{Moroshkin:2018}.

Since at the roton frequency the photon momentum is at many orders smaller than the roton momentum, the question arise: what is a mechanism to fullfil the momentum conservation law in the abovementioned processes? And why the narrow resonance is turned out to be observable upon the relative motion of the normal and superfluid components? All of that required new strong proofs and checks on the existence/absence of resonant photons exchange similar to that of interlevel transitions in atoms. We also posed a natural question on possible influence of an external electric field on the discovered the roton frequency narrow resonant line. Before, experimental studies of the Stark effect in helium were only conducted for the gas-state helium at high temperatures in the frequency range, where the fine structure of the excited levels of the helium atom becomes manifest \cite{Eliashevich:2001,LL:QM,Bonch-Bruevich:1968,Lamb:1957,Maiman:1957,Bethe:1957}. The present study is aimed at verification of the efficiency of an experimental device and at getting data on the impact of an external electric field on the resonant absorption line of the EM waves in the superfluid helium. 

The paper is organized as follows. In section II we discuss the experimental methodic. Here we recount the working parameters of our experimental setup and to overview the experimental conditions. In Section III we  recap the conditions under which the linear Stark effect in the liquid helium is revealed. Section IV contains the analysis and discussion of the results. Here we inspect the obtained result from different points of view, with trying to find the explanation of the Stark effect within various theoretical models and approaches. We review in brief the existent contradictions between theory and experiment and propose new mechanism of the electric dipole moment generation, closely related to having the relative flows of the normal and superfluid components in He II.  We summing up the results and sketching up possible lines of the development in the last section.

\section{Hallmarks of the experimental methodic}

The measuring device, mounted in a cylindric chamber of the diameter 80\,mm and of the length 100\,mm, consists of two parts: the system of radio measuring and the system of creating the hydrodynamic flows in the superfluid liquid (see Fig.1).

In this experiment we followed the previously used methodic of \cite{Rybalko:2007,Rybalko:2008,Rybalko:2008-1,Rybalko:2009,Rybalko:2010}. The usage of dielectric disk resonators (DDRs) gets advantaged in compare, e.g., to cylindric ones by the following: a) measuring the absorption of EM waves on the family of whispering gallery modes (WGMs) with a high Q factor allows us to involve a large range of frequencies upon appearing the circular relative flows of the normal and superfluid components; b) the artificially created relative motion of atoms compensates the Doppler shift upon exchanging among atoms the resonant photons of the roton energy (akin to the  M$\mathrm{\ddot{o}}$ssbauer effect) and makes possible to observe a narrow line; c) during the WGM damping time the interaction path of a running EM wave with the hydrodynamic superfluid flow is equivalent to the column of the liquid of a few hundred meters of length. Note to compare that in experiments on the Raman light scattering in the liquid helium the interaction path of an EM wave of an optical frequency was $\sim 0.1$\,m.

The disk dielectric resonator was made of the leuco sapphire; it has the diameter $20$\,mm and the width $1$\,mm. The resonator was embedded into the liquid helium and served as the main measuring element. The difference from the early used scheme consists in the following. To apply a homogeneous radial electric field, it was made a hole of the diameter 3 mm in the center of the resonator 1 (see Fig.1), where  a metal electrode 2 was inserted. The circular electrode 3~-- of the cylindric shape -- was installed coaxially to the resonator. It was made of a polyamide film of the width 30\,mkm, the diameter 22\,mm and of the height 0.91\,mm. The resonator and the circular electrode were supported by a frame; Fig.1 is not comprehensive in details. 
The thin conducting layer of graphite powder with the BF glue, electrically connected with grounding, was applied to the film surface. The value of constant electric field in the gap and its gradients was computed numerically. As follows from the numerics, the radial electric field in the gap between the cylindric surface of the resonator and the ring electrode 3 is homogeneous within 10 percent along all the cylinder height.\footnote{Computations on the dc-field homogenety in the gap were performed by A. A. Girich.} Since the capacitor formed by electrodes 2-3 has a complicated form and filled in part with the leuco sapphire, it is important to know the field value in the gap with the liquid. Its absolute value was calibrated in the following way: the conducting graphite film with the BF glue was put on the cylindrical resonator surface. The potential difference was applied in between the electrodes 2 and 3, and it was measured the potential between surfaces of the resonator and the electrode 3. After that the conducting layer has been washed away from the DDR surface.

The SHF radiation was entered the cryostat working zone, while its exit was performed by use of two rectangular waveguides 4 from the nickel silver. A microwave radiation was connected to the dielectric resonator (excitement and reception) by use of two pyramidal dielectric waveguides (antennas 5), located in the opposite directions on the disk plane. The ends of the antennas were covered with an absorber. Under the radiation supply via the waveguide it was excited a whispering gallery running wave, passing along the side surface (type HE$_{m,0,\d}$) of the resonator. The passing through the DDR signal was received by the reception antenna and registered by a semiconductor detector of the EM radiation, located at the opposite end of the nickel silver waveguide outside of the cryostat, i.e. at the room temperature (see \cite{Rybalko:2007} for more details).

Specific experiments created at the vacuum conditions of the working chamber with the DDR revealed that the placement of the conducting cylinder (the electrode 3 has the resistance of a few hundred M$\W$) on the 1mm distance from the resonator resulted in changing the resonant vibration frequency not more than 0.01\% in the range of frequencies higher than 100\,GHz. In particular, in the experiment at the frequency 62\,GHz the frequency shift was 5\,MHz, or 0.008\%, and the Q factor was fallen on 2\% that did not significantly impact on the character of EM waves absorption by the liquid. Upon the antennas location at the distance $\sim$2.0$\div$2.5\,mm from the DDR the coupling coefficient was changed not more than on 10\% that influenced only on the amplitude of the receiving signal.

It was applied a constant potential difference between the electrode 2 and the cylindric electrode 3, which created the gap electric field up to 40\,kV/m. Thus, the studied liquid in the gap between the cylindric surface of the resonator and the graphite film underwent the action of as the constant electric field $E_{dc}$ of the applied voltage, as well as an alternating electric field $E_{ac}$ created by the whispering gallery running wave. Both fields had the same radial direction from the center  of the resonator and the condition $E_{dc}\gg E_{ac}$ has been satisfied. 

The hydrodynamic system of superfluid flows creation contained two Kapitza heat guns 6, which are thermally insulated flasks connected by nozzles to a helium bath (working chamber). There were thermometers of the resistance 7 and the heaters 8 inside the flasks. This allowed us to superheat the liquid helium in a controlled way to $\sim $ 3\,mK inside the flask in respect to the bath temperature. When electric current is applied to the heater, from the nozzle 9 of the length $\sim$1\,cm and of the area $1000 \times 20$ mkm$^2$ it was flowed out a laminar jet of the normal component tangentially to the cylindric surface of the DDR. The normal component was slowing down at the resonator wall and in the gap, but excited circular flows of the superfluid component. Thus, in the experiments, by turning the different guns on, it was possible to create the superfluid component flows, the direction of which coincided or became opposite to the propagation direction of a whispering gallery wave. Forming a circular flow was accompanied with changing the top of the mode amplitude on 1$\div$4\%. Upon the heater current turning out the circular flows did not disappear for a long time. The artificially created circular flow could be destroyed (washed away) by a short time action of a mechanical pump 10, the jet of which is directed transversally to the gap flow (see Fig.1). It could also disappear due to the cryostat vibrations. The spectroscopy feature in presence of the superfluid flows in the gap was the appearance of the absorption narrow line at the roton frequency. Destroying the flow the narrow line disappeared. Results of fast scanning of the narrow resonance spectrum region after destroying the circular flow are presented on Fig.2. The signal/noise ratio was $S/N\sim 30$, so that the multiple frequency pass with a step of 8\,kHz/0.1\,sec allows us to observe the dynamics of appearing and narrowing the resonance line (registered curves 1,2,3,4,5 with the period of 3\,sec). Hence, it becomes clear that the circular flow was restored during the time $\sim$15\,sec. Further scanning was carried out with the step of 8\,kHz/1\,sec. The line became sharp. The large EM wave-liquid interaction path (300$\div$500\,m) together with the stability of the superfluid flows assist the narrow line detection. Unfortunately, we may only qualitatively observe the impact of the flows on the line, since studying the velocity field around the resonator quantitatively does not seem possible.

Within the experiments it has been registered the spectrum of the DDR modes and the narrow absorption lines of microwave radiation in He II on its background. The measurements were carried out at different power of the incoming radiation, at different directions and powers of heat flows, at different values of the constant electric field. The main result of these experiments consists in the confirmation of: 1) existence of circular flows around the DDR; 2) existence of the narrow line of the resonant absorption at the frequency, corresponding to the roton minimal energy $\D$ for a leuco sapphire resonator with the step of the WGM $\sim$1\,GHz. In experiments before it has been used the DDR of silica with the step between modes $\sim$2.3\,GHz. The basement width and its temperature dependence coincided with data on the roton line width \cite{Rybalko:2007,Rybalko:2008,Rybalko:2008-1,Rybalko:2009,Rybalko:2010}. These data are comparable in order with that of the early obtained in experiments on the neutron scattering \cite{Chagovets:2016pb,Rybalko:2017}. A wide roton line has also been observed in experiments on the Raman scattering of light by the liquid helium \cite{Chagovets:2017,Yayama:2018}.

\section{Splitting the absorption resonant line on applying a constant electric field}

As it was mentioned before, to observe the narrow line it was needed  to create a stationary superfluid flow in the gap. That is why at $T<T_\l$ the procedure of the spectrum measurement was preceded by switching the current through the located near the reception antenna gun heater on. Fig.3 includes preliminary results on observing the change of spectral characteristics of one of the whispering gallery modes $HE_{m,0,\d}$ with the azimuthal index $m=128$, the frequency $f\approx 180$\,GHz at the experimental conditions changing. This figure illustrates the change in the shape of the resonant curve of the mode as the temperature decreases from $T_\l$ to 1.4\,K. At the temperature $T=1.6$\,K (Fig.3a) the shape of the resonant curve does not have any peculiarities (the curve is approximated by the Lorentz (Peak shape) function) except that the ratio of ``liquid--vacuum'' state amplitudes in the working chamber with the DDR ($A_{\mathrm{ liquid}}/A_{\mathrm{vac}}$) becomes about 30\% larger of the similar relation for the modes with azimuthal indices $\D m=\pm 10$, that is in favor of a wide absorption basement at the roton frequencies region. Note, that at $T=1.6$\,K and at the normal conditions of the DDR's environment applying the constant electric field up to 40\,kV/m does not get a change in the resonant curve shape of this mode by its registration. 

Decreasing the temperature the mode location gets displaced in frequency due to changing the density of the liquid \cite{Smorodin:2017}; as a result, the mode and the narrow resonance absorption line mutually creep into each other. At the temperature $T=1.4$\,K the locations of the mode maximum and of the narrow resonance coincide (Fig.3b). Fig.3b corresponds to the case of null external constant electric field. In these conditions the resonant absorption frequency comes out as a dip (very narrow line) on the amplitude-frequency characteristic of the mode. Moreover, the width of the narrow line was influenced by heat flows created by Kapitsa's guns. So, it could be possible to select a power under which the width at the level of half of the amplitude was less than 40$\div$50\,kHz. 
Also it could be possible to initiate changing the regime of the resonant absorption to the induced radiation regime by use of the superfluid component flows along the direction of the whispering gallery wave propagation, similar to it was firstly demonstrated in \cite{Rybalko:2008}.

Below we will discuss on the narrow line in the absorption regime and on the influence on it the electric field. The scanning was carried out with the step $\sim$8\,kHz/1\,sec and the dc-electric field was turned on. Applying the electric field resulted in splitting the line on two. Figs.3c,d include spectra, registered at two values of the electric field. An almost symmetric picture of the lines displacement up and down in the frequency was observed, similar to that of the pioneering studies with the silica resonator \cite{Rybalko:2008ArXiv}. At the same time, the intensity of the lines was in 2$\div$2.5 times smaller than before turning the external field on. The signal of the resonant absorption narrow line was proportional to the power of the EM radiation of the generator, which does not exceed $10^{-3}$\,W.

On Fig.4 we present locations of frequencies for the left and the right resonances (the smaller and the larger  in frequencies $f_-$ and $f_+$) in dependence on the value of the constant electric field relatively to the narrow resonance frequency at the null field. As one can see, increasing the external field $E_{dc}$ the absorption frequencies change linearly with enlargement the field, and at large enough fields ($>20$\,kV/m) it is observed a deviation from the linear dependence towards lowering the frequencies. To clarify the nature of the observed results, we rearranged Figure 4 as follows: we put the data of the bottom curve on the horizontal axis (field axis). Then, the top curve data went the straight line in the whole range of frequencies. It became clear, that we experimentally observe the linear Stark effect. Unfortunately, we did not succeed with the simultaneous measurement of the line locations for fields greater than 40\,kV/m due to going the lines out the limits of resonant curve of the mode.

\section{Analysis and discussion of the results}

Previously, the experimental studies of the Stark effect have only been done for helium in the gas state at high temperatures \cite{Eliashevich:2001,LL:QM,Bonch-Bruevich:1968,Lamb:1957,Maiman:1957,Bethe:1957}. The experiments were performed at frequencies range where the thin structure of the helium atom levels come out, on the Balmer series of lines (see, e.g., \cite{Bonch-Bruevich:1968}). In that case, the Stark splitting of the excited states of the helium atom was studied. In all of the mentioned papers it was observed a quadratic dependence of the lines displacement from the value of the electric field. The field values were reached 10$^8$\,V/m.

In our experiments the field strength values were in a few orders less in the liquid and at $T<2$\,K. If the quadratic Stark effect would be observed, the top curve on Fig.4 should be deflected up of the linear one, while the bottom curve -- to the down. In fact, as the external field increases, both lines of the Stark spectrum are displaced in frequency down. One may suppose that a one-sided deviation from the linear law corresponds to a polarization of the liquid by an external field. Then the model, satisfying the observable set of phenomena, looks as follows: the liquid helium contains the interacting to each other particles with a dipole moment, which being in a local field, form energy states in it. Owing to the Doppler dispersion of energy of the individual particle states, a zone with the gap $\D$ is created in the liquid. Below $T_\l$ the dipole interaction becomes coherent. Laminar relative flows of two parts of the liquid promote the observation of a thin line of transitions.\footnote{In the alternate model of \cite{Rybalko:2009} it was pointed out that the roton is a collective excited mode of the helium as a many-particle system, hence one needs to take into account the change of state of the superfluid component in the momentum conservation law. The superfluid component, as a third body, could take the needed surplus of the momentum to itself to fulfill the conservation law; moreover, like a macroscopic object, practically without taking the energy away. This process is very like the same as that of the M$\mathrm{\ddot{o}}$ssbauer effect of transfering the momentum to a crystal as a whole. } The lines split upon the action of a constant external electric field.

\subsection{Analysis based on the known models and facts}

Within the model on the test the data obtained were processed (according to recommendations of \cite{Eliashevich:2001}, Chapter 5) by an empirical expression
\be
f_{\pm}=f_{rot}\pm\fr{dE_{dc}}{h}-\fr{\a E^2_{dc}}{2h}.
\la{fpm}
\ee
The first term on the r.h.s. of \rf{fpm} is the absorption frequency $f_{rot}$ in the null field on account of a compensation of the Doppler shift in frequencies by a relative motion of interacting particles. This frequency is related to the roton gap $\D$ by the relation $f_{rot}=\D/h=dE_{loc}/h$. The second term of \rf{fpm} describes the linear part of the Stark effect in the superfluid helium, and the third quadratic term corresponds to the polarization part. Here $h=6.63\cdot 10^{-34}$J$\cdot$sec is the Planck constant, $E_{dc}$ is the external field. The fitting parameters were the dipole moment of particles $d$, the local field $E_{loc}$, the polarization coefficient $\a$. The parameters $d$, $E_{loc}$, $\a$ were determined by experimental data and are in SI units as: $E_{loc}=(6.5\pm 1.5)\cdot 10^{11}$\,V/m, $d=(2.2\pm 0.7)\cdot 10^{-34}$\,C$\cdot$m, $\a=2\cdot 10^{-41}$ F$\cdot$m$^2$. The local field turned out to be comparable with the Coulomb field at an electron orbit, the dipole moment came at four orders less than that of polar molecules, the polarization coefficient turned out to be comparable with the polarization of helium atoms. 

A classical estimation of the Coulomb field value for the helium atom nucleus in the 1s state is about $0.5\div 1\cdot 10^{12}$\,V/m. If a possible error in the determination of the field value in the gap, coming from a thermic contraction of the polyamide film, would be taken into account, the charge displacement in a dipole falls into the range of $1\div 2\cdot 10^{-15}$\,m (less than the classical electron radius). From microscopic point of view the present understanding on the description of the $^4$He atom electron spectrum, being in a condensed matter of the same atoms, by use of 4 quantum numbers does not allow one to interpret the observed structure of the absorption at the range of 180\,GHz. But, apparently, upon the absorption and radiation the same mechanism as in an isolated atom takes place: upon transition from one state to another a system absorb or radiate the energy quanta.

It is believed that electrons of the liquid helium at low temperatures are of 1s-state, and the applied power of EM field $\sim 10^{-3}$\,W of the 180\,GHz frequency is not enough to excite more higher in energy levels. It is also believed that the helium atom in the 1s-state does not have the energy sublevels. However, the observed in the present work, as well as in \cite{Rybalko:2008ArXiv}, the linear Stark effect makes evident the existence of a particle with the electric dipole moment, that stimulates carrying out additional experiments on the sublevels existence or the presence of artifacts.

\subsection{Extension of currect theoretical models}

As we have noticed, the linear Stark effect in the liquid helium is highly likely explained by having the electric dipole moment in He II. Before turning to the discussion of possible mechanisms of its generation, recall that
studies in the electric dipole moment of elementary particles becomes more and more popular in the context of different extensions of the elementary particles Standard Model. For instance, the baryon-anti-baryon asymmetry of the visible Universe could be related to the CP-breaking, which, in its turn, is directly related to carrying the electric dipole moment by elementary particles. (See, e.g., \cite{Pospelov:2005,Engel:2013,Okawa:2019,Perez:2020,Chupp:2017} for  reviews, and the results of the ACME group \cite{Baron:2014,Baron:2017,Andreev:2018} in the measurements of the electron electric dipole moment (EDM) in atomic and molecular physics.)  Note, however, the principle mismatch between the results of ours and that of the ACME collaboration and theoretical computations based on QFT. The electron's electric dipole moment is theoretically predicted to be  $d_e\sim 1.6 \times 10^{-48}\div 10^{-51}$\,C$\cdot$m (see \cite{Okawa:2019} for recent computations of the $e^-$ EDM) and measured by the ACME collaboration to be $|d_e|<1.76\cdot 10^{-50}$\,C$\cdot$ m that (at least) 16 orders less that the electric dipole moment of the liquid helium. The value of the helion ($^3$He$^{++}$) light nuclei EDM is mostly determined by the neutron's EDM \cite{Chupp:2017}, and is of the order $d_{h}\sim 2.6 \cdot 10^{-47}$\,C$\cdot$\,m.
Therefore, the electric dipole moment of the liquid helium in presence of relative flows can not be ultimately treated as that of its elementary constituents, such as electrons, protons etc. (though we can not completely exclude such a possibility; see more on this in the last section), and has rather a collective excitation nature.

Various contemporary theoretical models of the liquid helium respond to a constant external electric field (see, e.g., \cite{Kruglov:2001,Loktev:2010,Loktev:2010ujp,Loktev:2011,Jain:2011,Tkachenko:2017,Loktev:2008,Mineev:2010,Khodusov:2012}; this is presumably a non-comprehensive list) divide on two principal classes, in dependence on chosen  consideration -- macroscopic or microscopic. The more preferable (more fundamental solution to the problem) microscopic point of view can not be succeeded in full extent due to a fundamental drawback in the liquid helium theory: it should be a theory of a strong-coupling system \cite{Sanders:2001,Schmitt:2015}. Moreover, within this picture the microscopic description has also to be extended to the roton by itself, making the ``What is a roton?'' question \cite{Donnelly:1997,Bedell:1982,Galli:1996,Reatto:1999,Harada:2009,Sanders:2001,Kruglov:2001,Tkachenko:2017,Loktev:2008,Loktev:2010} to be of extreme importance. 

The following rationals give a doubt in the models of the roton as a localized/bound state of atoms. Since the liquid helium is a quantum liquid, the laws of Quantum Mechanics have to be fully applicable to the case. The dispersion $|\d\vec{P}|$ in the momenta for the roton's localized state can not be greater than the roton energy minimum $\D$. In fact, as it follows from the roton energy-momentum relation (coming from the experimental $\E(\vec{P})$ curve),
\be
\E=\D+\fr{\left(\d\vec{P}\right)^2}{2\m}
\la{rotE}
\ee 
in the proper reference frame.
$\D \gg \left(\d\vec{P}\right)^2/2\m$, that, together with $\D \gg k_B T$, results in $\d P\sim \left(2\m k_B T  \right)^{1/2}$. (Here $\m \sim 0.16\, m_{\mathrm{He}}$.) Then, the Heisenberg uncertainty relation at $T \sim 1.4$\,K leads to 
\be
\d x \sim \fr{\hbar}{\d P}\sim 5.2\cdot 10^{-10}\,{\mathrm{m}},
\la{dx}
\ee
so that the dispersion in the coordinate becomes about twice larger than the natural interatomic distance in He II \cite{Donnelly:1998}. The roton, as a collection of quantum particles, becomes hard to localize (a non-dynamical, i.e., a topological and still uncovered conservation law is required to this end), and we conclude, as in before, that it is rather a collective excitation mode for the quantum liquid as a whole. 

Before turning to the macroscopic consideration let us make a brief conclusion on the microscopic atomic structure of He II in the superfluid phase. The analysis of the experimental data has resulted in fixing the parameters of \rf{fpm}; the intensity of external field is of $\sim 10^4$\,V/m (see Fig.\ref{fig:image4}). It is easy to see that the linear part of \rf{fpm} is about $6.6\cdot 10^{-30}$ in $h^{-1}\cdot$sec$^{-1}$. The quadratic in the field part of \rf{fpm} becomes $10^{-32}$ in the same units. Certainly, the linear part contributes the value of two orders higher than the quadratic part, resulting in the linear Stark effect. However, the linear Stark effect in atomic physics is rather an exclusion than a rule \cite{Bonch-Bruevich:1968,Kollath:1979,Windholz:2012}. In the case of the helium atoms \cite{Bethe:1957,Blinder:2004} it means that the most contribution follows from $1s1s\equiv 1s^2$ electron orbital configuration preferably than from $1s2s$ (or higher states like $1s2p$ etc.). In other words, the wave function of $^4$He atoms in the liquid state is mostly the product of two hydrogen-type ground state wave function: one of the doubled charge $Z=2$, and the other one of the charge $Z=1$. Hence, commonly one of the two helium electrons is located far enough of the atomic core, making possible the polarization by a weak external field.

Now we turn to phenomenology. The main advantage of the macroscopic description (cf. for instance \cite{Donnelly:1997, Sanders:2001, Schmitt:2015}) consists in the transition to the dual description of the quantum liquid at the strong-coupling regime in terms of weakly-interacting quasi-particles -- phonons and rotons (last time maxons were separated to additional class of quasi-particles). Therefore, the microscopic structure of the quantum liquid constituents becomes not so important, so that the description is based on average characteristics of the liquid.   

One of the characteristics of He II is the frequency corresponding to the roton minimum, $f_{rot}\sim$\,180\,GHz, entering the empirical expression \rf{fpm} in before. Manifestation of this quantity in many experiments on the electrical activity of the liquid helium (see, e.g., \cite{Rybalko:2004,Rybalko:2005,Rybalko:2007,Rybalko:2008,Rybalko:2008-1,Yayama:2018}) sets possible to conclude that this quantity is the internal resonant frequency of the medium and is that of fundamental importance.

Within the quasi-particles formalism 
the narrow dip near the resonant frequency of He II in a weak external electric field can be treated as a corollary of dynamical processes of absorption/radiation of a super-high frequency wave by the medium, or, more precisely, of Quantum Electrodynamics of rotons. For instance, the analysis of the energy and momentum conservation laws for the Raman scattering of the light \cite{Khodusov:2012} results in a narrow domain of physically admissible frequencies (of the kHz order, cf. Fig.\ref{fig:image3}) and of wave vectors of the EM radiation near the roton minimum. It is worth mentioning the admissible narrow domain in the wave vectors of radiation includes momenta near the roton minimum which are opposite in the direction (resulted in Fig. 3c or Fig. 3d). It corresponds, by the momentum conservation law, to 2$\ra$2 processes of photons interacting with rotons,\footnote{The photon's $C$-parity excludes processes with a single photon, while kinematical arguments prohibit the appearance of a single roton in the Raman scattering.} the momenta of which are of the same order, but opposite in the direction~-- the so-called $R^+$ and $R^-$ rotons \cite{Donnelly:1997}, in accordance to the orientation of their group velocity with respect to the direction of the heat flow (towards or against). 

Since the electric respond of the liquid directly depends on having the heat flows, to which $R^+$/$R^-$ rotons inherent, the induced electric dipole moment has naturally to be dependent on main characteristics of the flow -- velocity and/or acceleration -- and includes $f_{rot}$ as a fundamental characteristic of the quantum liquid respond. On account of these observation the dimensionality arguments suggest the following expression for the induced electric dipole moment:
\be
\vec{d}=\fr{ e\hbar}{\D}\, \vec{v}_g=\fr{e \hbar}{\D} \left(\fr{P-P_0}{\m}\right)\fr{\vec{P}}{P},
\la{ddef}
\ee
with the roton's group velocity $\vec{v}_g$. As a result, standard computations of the average value of $\vec{d}$ with the equilibrium distribution function of the rotons (details of which we will postpone for publishing elsewhere) and with the known values of He II parameters (see, e.g., \cite{Donnelly:1998}), lead to the following value of $\bar{d}\equiv \overline{|\vec{d}\,|}$ at $T\sim 1.4$\,K: 
\be
\bar{d} \approx 2.06 \cdot 10^{-34}\,{\mathrm{C \cdot m}},
\la{daver}
\ee
which is close to the experimentally observed value $(2.2\pm 0.7)\cdot 10^{-34}$\,C$\cdot$m.\footnote{Note that the right order of the induced electric dipole moment has been recovered from other phenomenological rationales in \cite{Mineev:2010}. However, in contrast to the approach of \cite{Mineev:2010}, where at least two arbitrary fitting parameters were used, the proposed expression of the induced electric dipole moment \rf{ddef} contains solely the proper characteristics of the liquid helium. }

\section{Summary and conclusions}

We have demonstrated the device that makes possible to excite the circular superfluid flows along a running wave of the whispering gallery of the dielectric disk resonator and to carry out the electromagnetic spectroscopy of He II. It has been revealed that at the laminar relative flows of the normal and of the superfluid components it is observed the narrow line of the EM spectrum at the roton frequency (180\,GHz), which splits by a constant electric field.

The observed linear Stark effect is directly related to the induced electric dipole moment (EDM) of the liquid helium. However, our arguments support the view of the electric dipole moment as a reaction of the entire medium to a given action of a weak external field, rather than a contribution of its separate microscopic constituents to the effect. One of the facts making evident this point of view is given by comparing the obtained value of the liquid helium EDM to the electron's electric dipole moment: the value of the latter, measured by the ACME group, is in 16 orders less that the former. Possible EDM values of the helion -- the doubly ionized $^3$He atom -- are of 13 orders less than that of the liquid helium. In theory, the value of particle density in the liquid helium is large enough to fill this huge gap. However, the specific realization of a mechanism leading to exact matching the average value of the total electric dipole moment of the He II constituents to that of our findings definitely deserves additional studies.  Hence, at the present state of the art, we conclude that correct microscopic arguments, based on the internal (atomic-electron) structure of the considered Bose-Einstein Condensate (BEC), are either still missing to reproduce the precise value of the He II EDM, or can not be applied in full extent due to the strong-coupling nature of the liquid. At the same time the consideration of the collective behavior of the BEC as a whole (in terms of quasi-particles) is quite reasonable and is one of the ways to describe different non-apparent properties of the liquid helium.

Surely, a lot of work should be done before achieving the ultimate explanation of the observed phenomena discussed here. From the point of view of theory, the proposed by us expression for the EDM (formula \rf{ddef}) has to be refined for a non-trivial geometry of the experiment (such as \cite{Rybalko:2005}); from the experimental point of view it would be interesting to find frequencies of the liquid respond corresponding other (than the Raman scattering) admissible processes of interacting EM waves with collective excitations of He II. We are planning to turn to these and other related tasks in the future.

We conclude with noting the following important detail. Specifically, the EDM of He II is that of its collective excitation modes -- quasi-particles. It was pointed out long before \cite{Iwamoto:1989} the collective excitations can also be supplied with different quantum numbers, usually inherent to elementary particles. One may assign to quasi-particles an angular momentum, analog of spin, endow them with the parity quantum number and so on. Our findings extend this list to the electric dipole moment, which now can be associated to rotons. So we can observe, again and again, a mystical intertwining between Elementary Particle and Condensed Matter Physics.

\section*{Acknowledgements} We thank M. I. Nakhimovich for making the dielectric disk resonator and dielectric antennas, and A. A. Girich for computations of field in the gap in between the resonator and antennas. 

\newpage

\begin{figure}
\center{\includegraphics[width=0.6\linewidth]{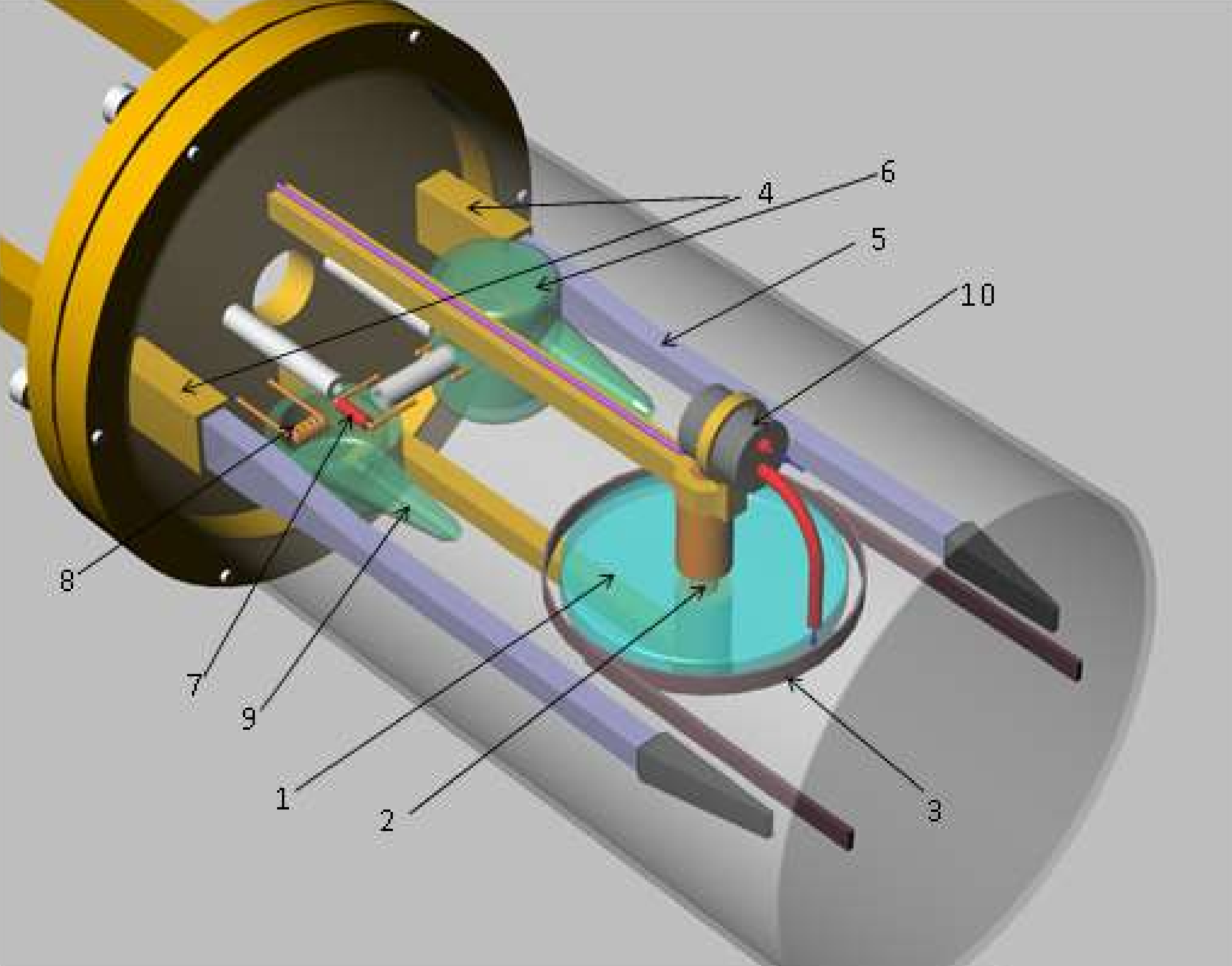}}
\caption{Schematic diagram of creating circular superfluid flows to obsere a narrow line and its splitting by a constant electric field. The notation is: ${\bf 1}$ is the dielectric disk resonator; ${\bf 2}$ is the dielectric support of the resonator with a metallic electrode inside; ${\bf 3}$ is the polyamide cylinder with a conducing film; ${\bf 4}$ are the waveguides of nickel silver; ${\bf 5}$ are the dielectric antennas; ${\bf 6}$ are the thermo-isolated flasks; ${\bf 7}$ is the resistance thermometer RuO$_2$; ${\bf 8}$ is the heater; ${\bf 9}$ is the nozzle; ${\bf 10}$ is the mechanical vibrational pump to create an axial flow of the liquid in the gap.   }
\label{fig:image1}
\end{figure}

\bsk\bsk
\begin{figure}
\center{\includegraphics[width=0.99\linewidth]{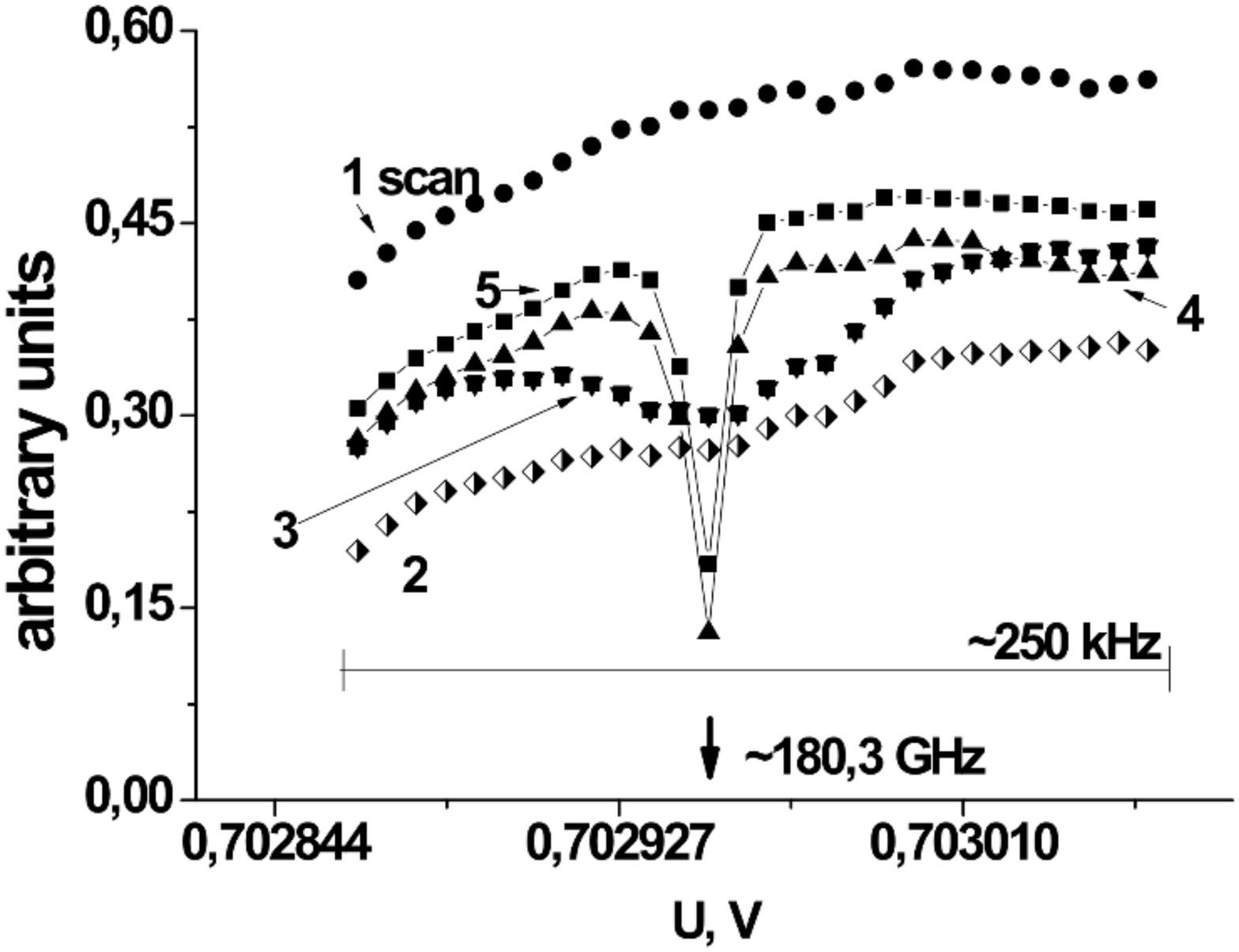}}
\caption{The amplitude-frequency dependence upon scanning the region of the narrow line of the spectrum in arbitrary units. The horizontal axis corresponds to the generator tuning voltage in frequency, the value of which was increased by a surge in every 0.1 second. The Lock-in time constant is also 0.1 sec. Every 3 seconds the scanning process was returned to the starting point. The first scanning corresponds to the point in time after the turning the mechanical pump off. The next scans reflect the dynamics of recovering the narrow line in the spectrum.}
\label{fig:image2}
\end{figure}

\bsk\bsk
\begin{figure}[h]
\begin{minipage}[h]{0.4\linewidth}
\center{\includegraphics[width=1.45\linewidth]{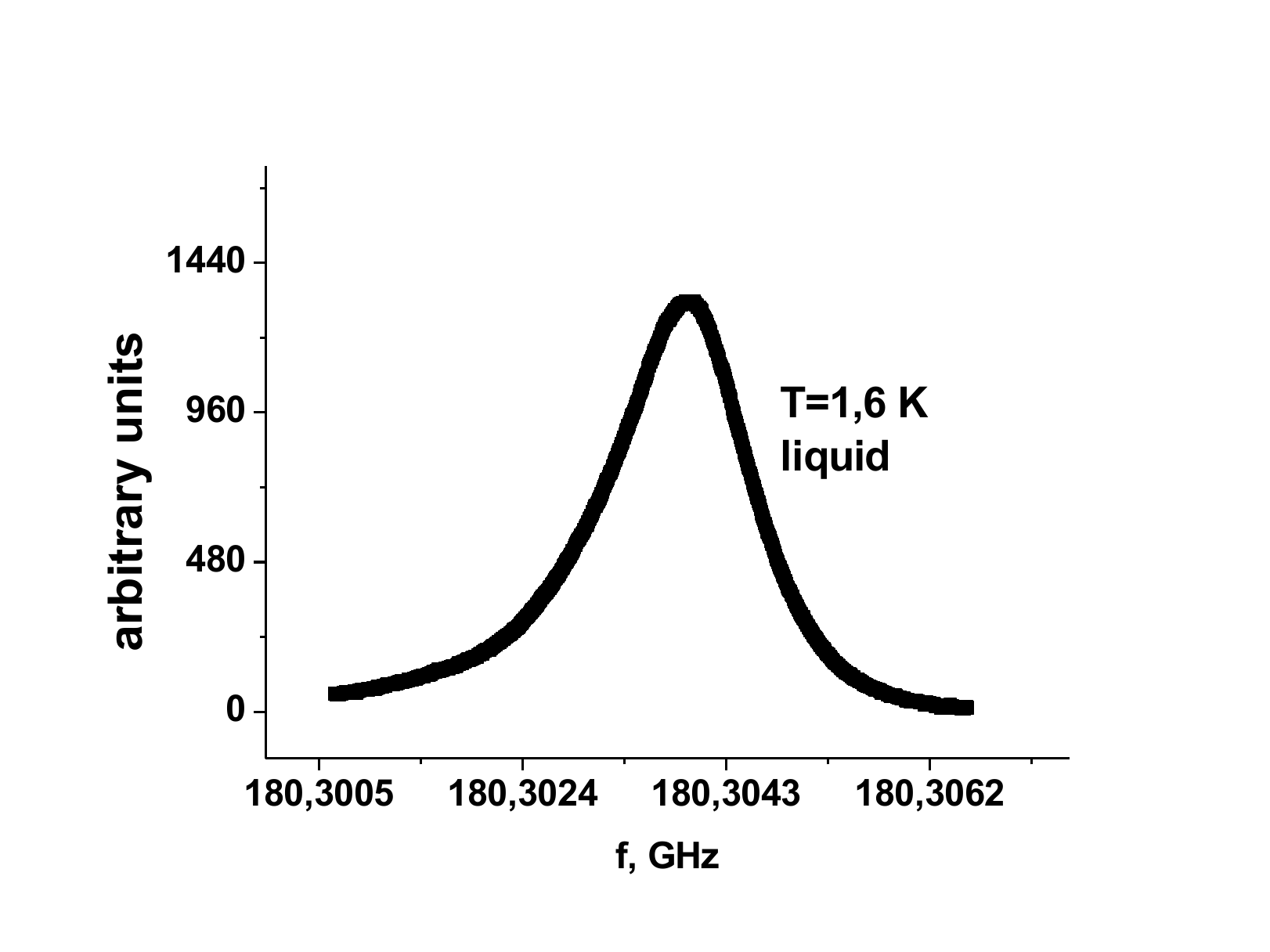} \\ a)}
\end{minipage}
\hfill
\begin{minipage}[h]{0.4\linewidth}
\center{\includegraphics[width=1.2\linewidth]{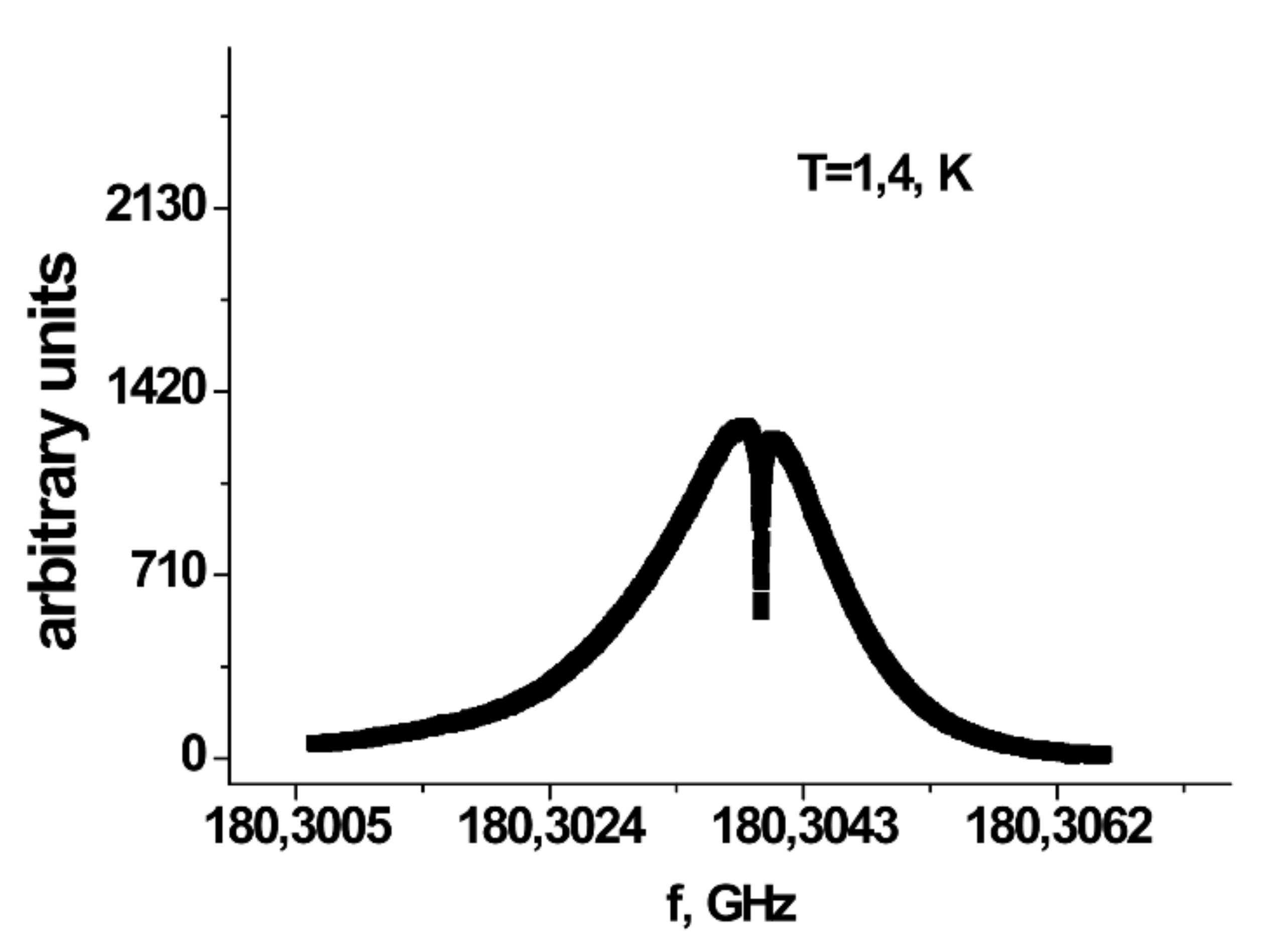} \\ b)}
\end{minipage}
\hfill
\begin{minipage}[h]{0.4\linewidth}
\center{\includegraphics[width=1.2\linewidth]{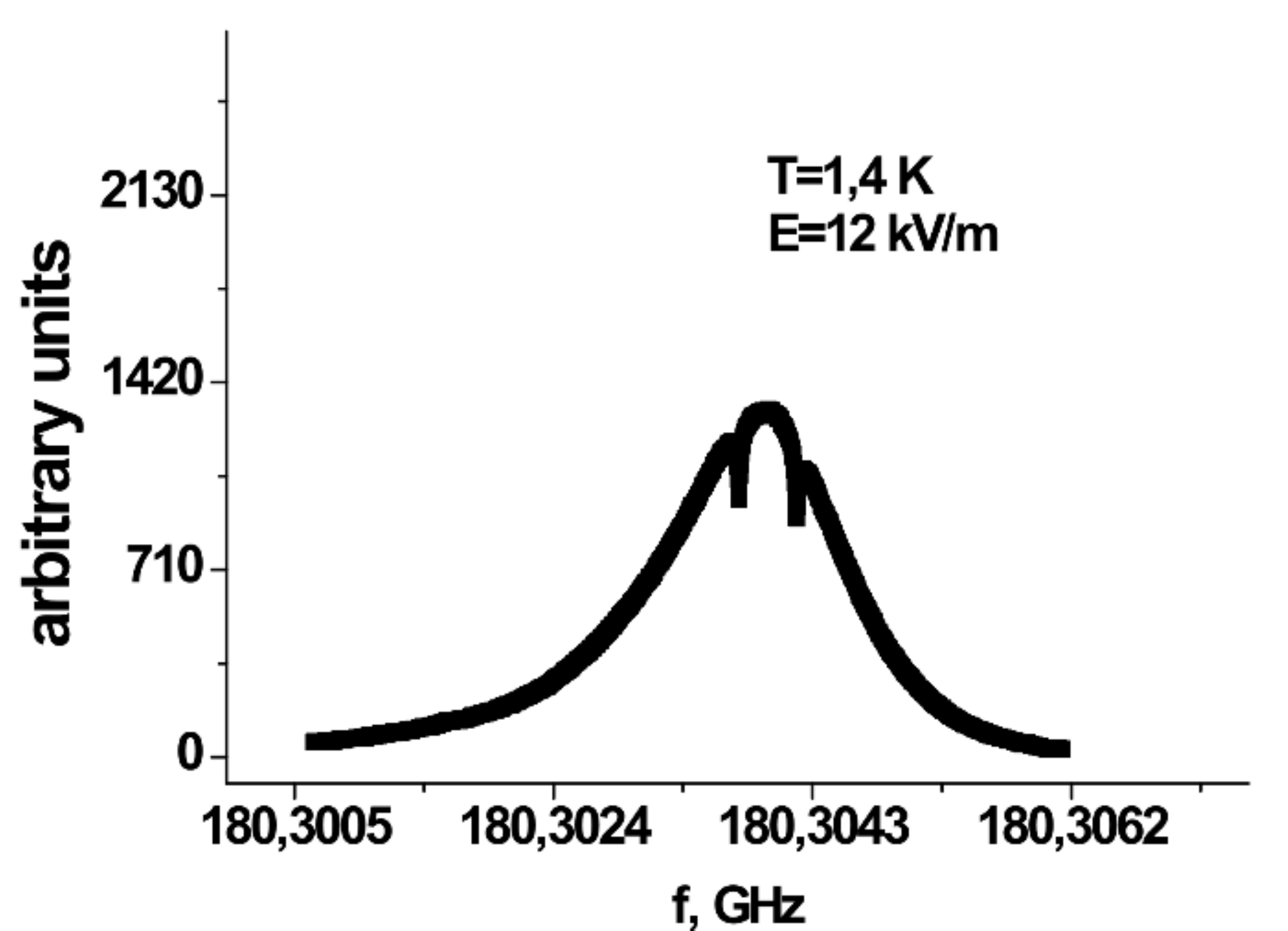} \\ c)}
\end{minipage}
\hfill
\begin{minipage}[h]{0.4\linewidth}
\center{\includegraphics[width=1.2\linewidth]{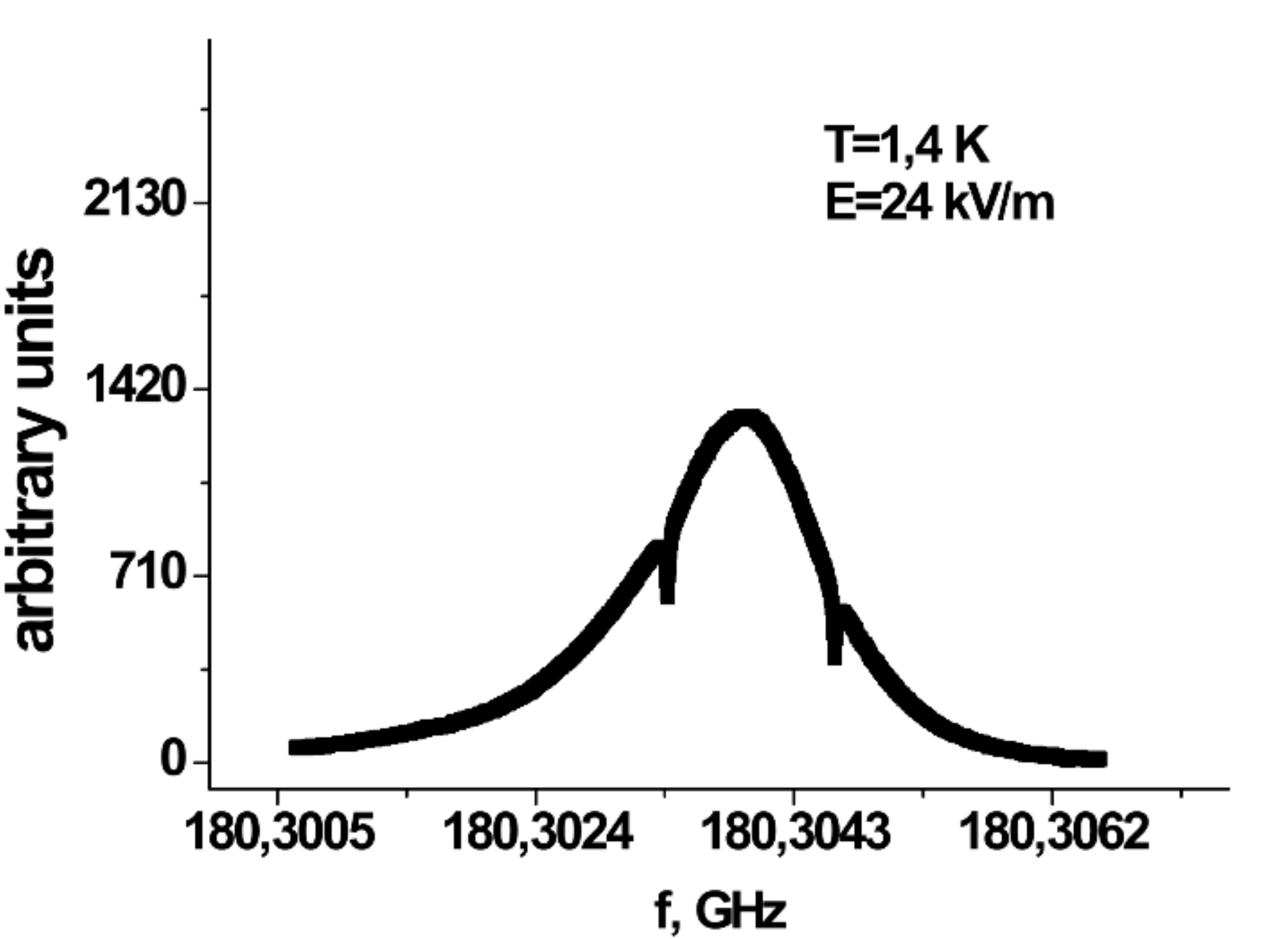} \\ d)}
\end{minipage}
\caption{The amplitude-frequency characteristics of the DDR mode at different temperatures. The Lock-in time constant is 1 sec, the step change in frequency is $\sim 8$kHz. See Section III for more details.}
\label{fig:image3}
\end{figure}


\bsk\bsk
\begin{figure}
\center{\includegraphics[width=1.0\linewidth]{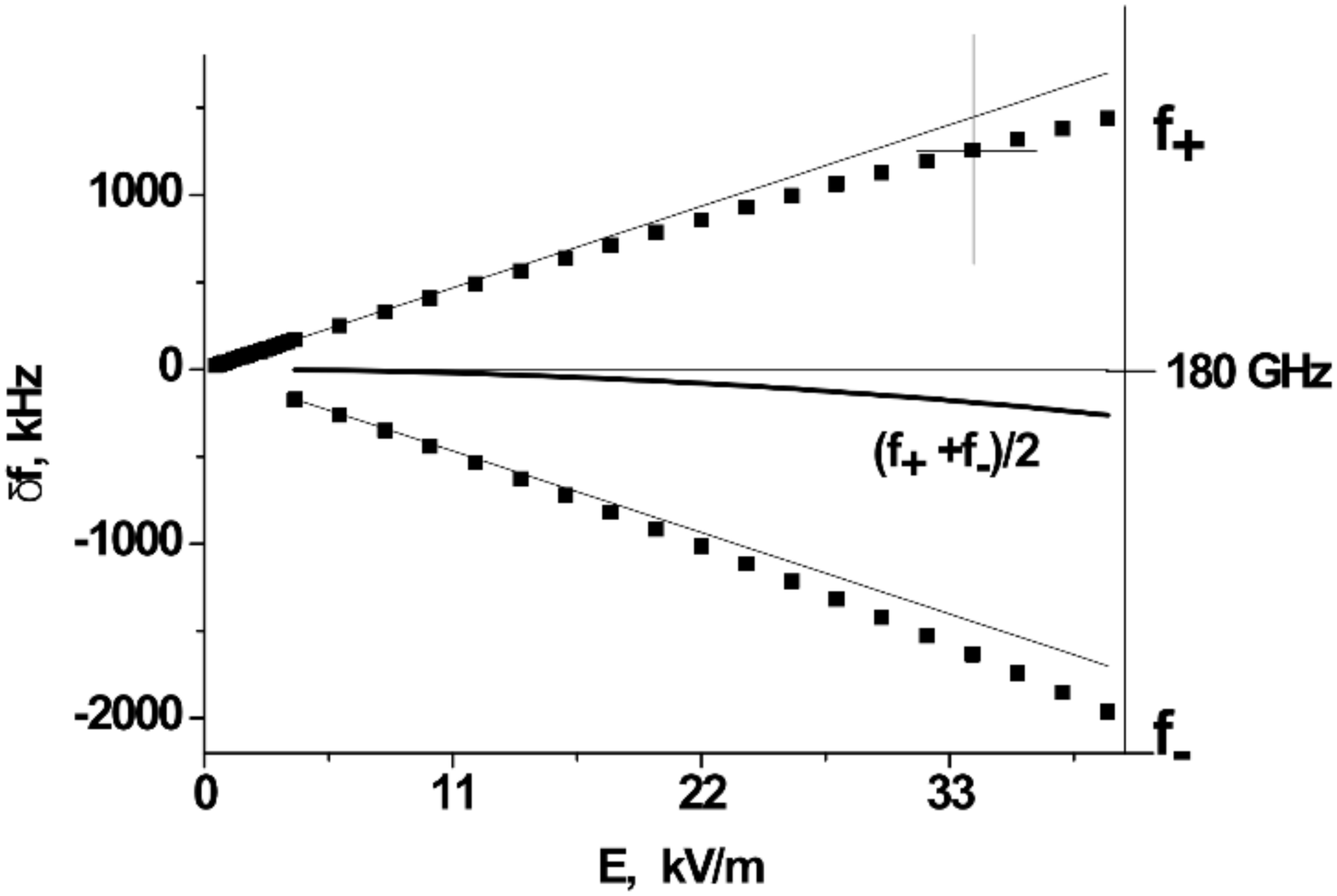}}
\caption{The dependence of the splitting frequency of the narrow line on the electric field value. Points at $E<5$kV/sm are borrowed from [36]. The error bars point the error values in the determination of absolute values of the electric field in the gap and of the 180GHz frequency.}
\label{fig:image4}
\end{figure}



\end{document}